\documentclass[12pt]{article}
 \usepackage{epsfig}
 \def\be{\begin{equation}}
 \def\ee{\end{equation}}
 \def\bea{\begin{eqnarray}}
 \def\eea{\end{eqnarray}}
 \usepackage{graphicx}

 \catcode`\@=11
 \def\lsim{\mathrel{\mathpalette\@versim<}}
 \def\gsim{\mathrel{\mathpalette\@versim>}}
 \def\@versim#1#2{\vcenter{\offinterlineskip
 \ialign{$\m@th#1\hfil##\hfil$\crcr#2\crcr\sim\crcr } }}
 \catcode`\@=12

 \catcode`@=12
\begin{document}
 \thispagestyle{empty}
 \begin{flushright}
 UCRHEP-T591\\
 Sep 2018\
 \end{flushright}
 \vspace{0.6in}
 \begin{center}
 {\LARGE \bf Leptonic Dark Matter with \\Scalar  
Dilepton Mediator\\}
 \vspace{1.2in}
 {\bf Ernest Ma\\}
 \vspace{0.2in}
{\sl Physics and Astronomy Department,\\ 
University of California, Riverside, California 92521, USA\\}
\vspace{0.1in}
 \end{center}
 \vspace{1.2in}

\begin{abstract}\
A simple and elegant mechanism is proposed to resolve the problem of having a 
light scalar mediator for self-interacting dark matter and the resulting 
disruption to the cosmic microwave background (CMB) at late times by the 
former's enhanced Sommerfeld production and decay.  The crucial idea is 
to have Dirac neutrinos with the conservation of U(1) or $Z_N$ lepton 
number extended to the dark sector.  The simplest scenario consists 
of scalar or fermion dark matter with unit lepton number accompanied 
by a light scalar dilepton mediator, which decays to two neutrinos.
\end{abstract}

\newpage
\baselineskip 24pt
\noindent \underline{\it Introduction}~:~
Whereas the existence of dark matter is universally accepted, its nature is 
unknown~\cite{y17}.  Even if it is assumed that it consists of one specific 
fundamental particle, there is no information of what properties it must 
have, such as mass, spin, and possible interactions with itself or other 
particles. It must of course have the correct relic abundance and satisfy 
all present experimental bounds, but these constraints are not very 
restrictive and are easily satisfied by many models.  This uncertain state 
of affairs has led to numerous diverse proposals for specific candidates 
of dark matter, most of which are at present not excluded.  Is there a hint 
which would allow us to zero in on one specific candidate?  In this paper, 
it is argued that this may indeed be the case if dark matter is assumed to 
have self-interactions mediated by a light scalar~\cite{kkpy17} to explain 
the central flatness of the density profile of dwarf galaxies~\cite{detal09}.  
The key is that this light scalar mediator has a large production cross 
section through Sommerfeld enhancement~\cite{s31} at late times, and its 
decay to electrons and photons would disrupt~\cite{gibm09} the 
cosmic microwave background (CMB) and be ruled out~\cite{bksw17} by the 
precise observation data now available~\cite{planck16}.  To escape this 
conundrum, a simple and elegant mechanism is proposed, which points to 
a specific scenario of dark matter as detailed below.

\noindent \underline{\it Predestined Leptonic Scalar Dark Matter}~:~
Consider the simple extension of the standard model (SM) of quarks and 
leptons with only three singlet right-handed neutrinos $\nu_R$, but 
also with the conservation of $U(1)$ or $Z_N$ lepton number.  The theoretical 
questions of where and how this symmetry may come from, and why the masses 
are so small, will be discussed later.  In this framework, neutrinos are 
assumed Dirac fermions with small masses which could be natural consequences 
of various known mechanisms~\cite{bmpv16,mp17,ms18}.

A simple and elegant scenario of self-interacting dark matter is now 
possible with the addition of a neutral scalar ($\chi$) having $L=1$ 
and one other ($\zeta$) having $L=2$.  Note first that $\chi$, being a 
scalar, is now absolutely stable from the conservation of $U(1)_L$ lepton 
number alone, or from $Z_N$ lepton number with $N \neq 2$.  It is the 
analog of deriving~\cite{m15} dark parity from lepton parity using 
$(-1)^{L+2j}$ in the case of Majorana neutrinos $(N =2)$.  Thus 
$\chi$ may be considered predestined dark matter~\cite{m18} because it 
is the automatic result of an existing symmetry and its chosen particle 
content.

By itself, $\chi$ acts as a dark-matter candidate in much the same way as 
the simplest model of dark matter using a real scalar of odd dark 
parity~\cite{gambit17}. 
With the scalar dilepton $\zeta$, it may now have the interaction 
$\zeta^* \chi^2$ which enables the enhanced scattering of 
$\chi \chi^* \to \chi^* \chi$ through the exchange of $\zeta$ as a 
light scalar mediator.  The problem of disrupting the CMB 
is solved because $\zeta$ only decays to two neutrinos through the allowed 
$\zeta^* \nu_R \nu_R$ coupling.  This works because $\nu_R$ combines with 
$\nu_L$ to form a Dirac neutrino of very small mass.  In the canonical 
seesaw mechanism of small Majorana neutrino masses, $U(1)_L$ breaks 
to lepton parity, and $\nu_R$ has its own very large Majorana mass.  
The resulting $\nu_L - \nu_R$ mixing is very small, so the decay lifetime 
of $\zeta \to \nu_L \nu_L$ through this mixing is very long~\cite{mpsz15} 
if it were the only decay mode.  However, since only lepton parity is 
conserved in that scenario, $\zeta$ is even and mixes with the SM Higgs 
boson and could also decay through that mixing to electrons and photons, 
thereby disrupting the CMB, as in an earlier proposal~\cite{m17}.  
If the light scalar mediator decays 
dominantly to $\nu_L \nu_L$, then it must belong mostly to an 
electroweak triplet.  A model of this kind is possible 
but the details are much more complicated~\cite{mm17}.

It is important to note that with neutrinos as Dirac fermions and the 
$\zeta^* \nu_R \nu_R$ interaction, the two other astrophysical anomalies 
regarding dark matter, i.e. the missing-satellite and too-big-to-fail 
problems, may also be solved, using the drag on dark matter by the 
cosmic neutrino background, as pointed out in 
Refs.~\cite{betal14,setal15,setal16}.  The 
self-interaction of $\nu_R$ through $\zeta$ may also be used in 
explaining dark energy as neutrino condensates~\cite{bdmrs10}, whereas 
the Dirac nature of neutrinos is known to be a possible origin of 
the baryon asymmetry of the Universe through neutrinogenesis~\cite{dlrw00}.

\noindent \underline{\it Details of the Dark Sector}~:~
The most general scalar potential consisting of $\chi$, $\zeta$, and the 
SM Higgs doublet $\Phi$ is given by
\begin{eqnarray}
V &=& \mu_0^2 \Phi^\dagger \Phi + \mu_1^2 \chi^* \chi + \mu_2^2 \zeta^* \zeta 
+ [\mu_{12} \zeta^* \chi^2 + H.c.] \nonumber \\ 
&+& {1 \over 2} \lambda_0 (\Phi^\dagger \Phi)^2 + {1 \over 2} \lambda_1 
(\chi^* \chi)^2 + {1 \over 2} \lambda_2 (\zeta^* \zeta)^2 \nonumber \\ 
&+& \lambda_{01} (\Phi^\dagger \Phi)(\chi^* \chi) + \lambda_{02} 
(\Phi^\dagger \Phi)(\zeta^* \zeta) + \lambda_{12} (\chi^* \chi)
(\zeta^* \zeta). 
\end{eqnarray}
With $\langle \phi^0 \rangle = v = 174$ GeV,
\begin{equation}
m^2_H = 2 \lambda_0 v^2 = (125~{\rm GeV})^2, ~~~ m^2_\chi = \mu_1^2 + 
\lambda_{01} v^2, ~~~ m^2_\zeta = \mu_2^2 + \lambda_{02} v^2.
\end{equation}
The only new Yukawa couplings are 
\begin{equation}
{\cal L}_Y = f_{ij} \zeta^* \nu_{Ri} \nu_{Rj} + H.c.
\end{equation}
With the assumption of $U(1)_L$ conservation, the $f_{ij}$ terms fix the 
leptonic charge of $\zeta$ to be 2, and the $\mu_{12}$ term fixes the 
leptonic charge of $\chi$ to be 1.  Since $\Phi$ has $L=0$, there is no 
mixing among $H$, $\chi$, and $\zeta$.  In this simplest model, 
$\chi$ is dark matter, whose stability depends only on conserved lepton 
number, and $\zeta$ is its light dilepton mediator.

The elastic scattering of $\chi \chi^* \to \chi^* \chi$ proceeds via the 
diagram of Fig.~1.
\begin{figure}[htb]
\vspace*{-6cm}
\hspace*{-3cm}
\includegraphics[scale=1.0]{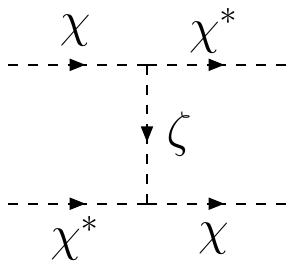}
\vspace*{-21.5cm}
\caption{Dark matter $\chi$ elastic scattering by exchanging $\zeta$.}
\end{figure}
The resulting cross section is given by
\begin{equation}
\sigma_{el} = {\mu^4_{12} \over 4 \pi m^2_{\chi} m^4_\zeta}.
\end{equation}
Note that for a heavy $\chi$ and a light $\zeta$, this may be enhanced 
sufficiently to affect the central density profile of the dark matter 
distribution, whereas the $s$-channel processes 
$\chi \chi^* \to H \to \chi \chi^*$ and $\chi \chi \to \zeta^* \to \chi \chi$ 
as well as the $t$-channel processes via $H$ exchange are negligible compared 
to it.  The same applies to the $\lambda_1$ quartic interaction term.  

The relic abundance of $\chi$ is determined by its annihilation cross section 
$\times$ relative velocity as the temperature of the Universe falls below its 
mass.  Assuming that $\lambda_{01}$ is negligible to avoid the constraint 
from $H$ exchange in $\chi$ elastic scattering off nuclei in underground 
direct-search experiments, the main contributions of 
$\chi \chi^* \to \zeta \zeta^*$ come from the $\lambda_{12}$ 
quartic interaction and $\mu_{12}$, i.e.
\begin{equation}
\sigma_{ann} v_{rel} = {1 \over 32 \pi m^2_\chi} \left( \lambda_{12} - 
{2 \mu^2_{12} \over m^2_\chi} \right)^2.
\end{equation}
At temperatures above $m_\chi$, the $\zeta \zeta^* H$ interaction with 
strength $\sqrt{2} v \lambda_{02}$ allows $\zeta$ (and thus $\chi$) to be 
in thermal equilibrium with the SM particles.  After $\chi$ freezes out,  
$\zeta$ eventually decays to neutrinos via the $f_{ij}$ terms with 
a decay rate given by
\begin{equation}
\Gamma(\zeta \to \nu_R \nu_R) = {m_\zeta \sum_{i,j} |f_{ij}|^2 \over 4 \pi}.
\end{equation}

As a numerical example, let $m_\chi = 150$ GeV, then 
$\sigma_{ann} v_{rel} = 4.4 \times 10^{-26}~cm^3/s$ is obtained with 
$\lambda_{12} - 2\mu^2_{12}/m_\chi^2 = \pm 0.0923$, which implies that 
$|\mu_{12}| < 32.2$ GeV if $\lambda_{12} < 0$.  Setting $\sigma_{el}/m_\chi$ 
equal to the benchmark value of 1 $cm^2/g$ for self-interacting dark 
matter, the ratio $m_\zeta/|\mu_{12}| = 0.0015$ is required.  Hence 
$m_\zeta < 48.3$ MeV if $\lambda_{12} < 0$.   Using $m_\zeta = 40$ MeV, 
its decay lifetime is about $2 \times 10^{-16} ~s$ for 
$\sum_{i,j} |f_{ij}|^2 = 10^{-6}$.  This means that before the onset of 
big bang nucleosynthesis (BBN), $\zeta$ has all decayed away and 
$\nu_R$ decouples from the rest of the SM particles.  In particular, 
the number of effective massless degrees of freedom used in the 
standard BBN scenario is unchanged. In Fig.~2 the allowed range  
$10 < m_\zeta < 100$ MeV is shown against $100 < m_\chi < 200$ GeV for 
$\lambda_{12} = 0, \pm 0.1$.

\begin{figure}
\hspace*{-1cm}
\includegraphics[scale=1.0]{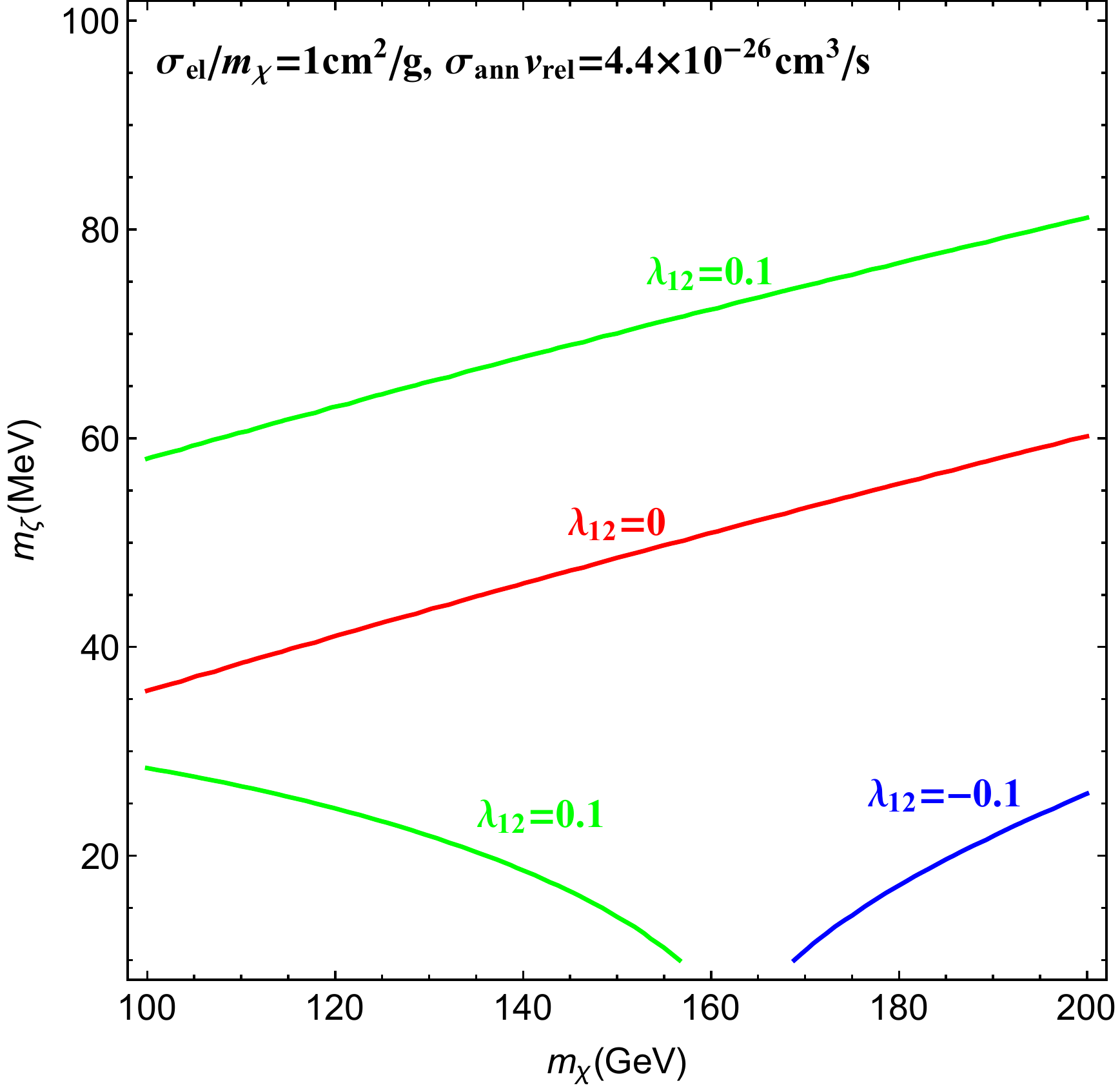}
\vspace*{-0.5cm}
\caption{Allowed region in $m_\zeta-m_\chi$ space for self-interacting dark 
matter.}
\end{figure}

The interaction of the leptonic dark matter $\chi$ with quarks and leptons 
is through the SM Higgs boson $H$.  For $m_\chi = 150$ GeV, this elastic 
cross section $\sigma_0$ is bounded by the latest experimental 
result~\cite{xenon17} to be below $2 \times 10^{-46}~cm^2$.  This 
translates to an upper bound~\cite{kmppz18} of $4.4 \times 10^{-4}$ 
for the $\lambda_{01}$ quartic coupling of Eq.~(1).
As for the $\lambda_{02}$ quartic coupling, it is constrained by the 
invisible decay width of $H \to \zeta \zeta^*$, i.e.
\begin{equation}
\Gamma (H \to \zeta \zeta^*) = {\lambda_{02}^2 v^2 \over 4 \pi m_H}.
\end{equation}
Assuming that this is less than 10\% of the SM width of 4.12 MeV, 
then $\lambda_{02} < 4.6 \times 10^{-3}$.

\noindent \underline{\it Leptonic Fermion Dark Matter}~:~
The scalar leptonic dark matter $\chi$ may be replaced by a Dirac fermion 
$\psi$.  Now the singlets $\psi_R$ and $\nu_R$ both have $L=1$ and must 
be distinguished.  In other words, a dark $Z_2$ parity must be imposed, so 
that $\psi$ is odd and all other particles even.  The scalar sector 
consists only of $\Phi$ and $\zeta$.  The Yukawa sector has the new 
terms
\begin{equation}
{\cal L}_Y = f_L \zeta^* \psi_L \psi_L + f_R \zeta^* \psi_R \psi_R + H.c.
\end{equation}
The analog of Fig.~1 is then Fig.~3.
\begin{figure}[htb]
\vspace*{-6cm}
\hspace*{-3cm}
\includegraphics[scale=1.0]{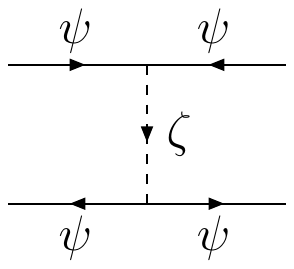}
\vspace*{-21.5cm}
\caption{Dark matter $\psi$ elastic scattering by exchanging $\zeta$.}
\end{figure}
The resulting cross section is given by
\begin{equation}
\sigma_{el} = {(f_L + f_R)^4 m^2_\psi \over 4 \pi m^4_\zeta}.
\end{equation}
The analog of Eq.~(5) is
\begin{equation}
\sigma_{ann} v_{rel} = {f_L^2 f_R^2 \over \pi m^2_\psi}.
\end{equation}
Using again $m_\psi = 150$ GeV as an example, and assuming $f_L=f_R=f$, 
then $f = 0.128$ is obtained, as well as $m_\zeta = 58$ MeV.

Regarding direct detection, $\psi$ has no Yukawa coupling with the 
SM Higgs boson at tree level, but may do so in one loop as shown in Fig.~4.
\begin{figure}[htb]
\vspace*{-5cm}
\hspace*{-3cm}
\includegraphics[scale=1.0]{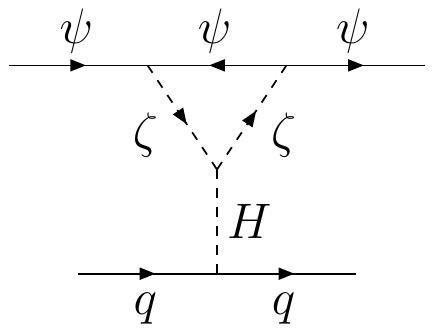}
\vspace*{-21.5cm}
\caption{Dark matter $\psi$ scattering off nuclei through $H$ exchange.}
\end{figure}
However, since $\lambda_{02} < 4.6 \times 10^{-3}$ and $f_L f_R = 0.0164$, 
this contribution is very much negligible.

\noindent \underline{\it Predestined Leptonic Fermion Dark Matter}~:~
To eliminate the need to impose a dark parity for leptonic fermion 
dark matter, $\psi$ may be assigned $L=2$ with the addition of a scalar 
$\eta$ with $L=4$.  Then $\psi$ is predestined dark matter~\cite{m18} 
because it is stable as the result of an existing symmetry and its 
particle content.  Now $\eta$ (instead of $\zeta$) acts as its light 
mediator, with $\eta$ decaying into $\zeta \zeta$, then 
$\zeta \to \nu_R \nu_R$.

\noindent \underline{\it Gauge Origin of Lepton Number Conservation}~:~
It is well-known that with three singlet right-handed neutrinos, $B-L$ 
may be implemented as an anomaly-free gauge symmetry.  The conventional 
approach is to break gauge $B-L$ spontaneously with a scalar having $L=2$. 
Since this scalar also couples to $\nu_R \nu_R$, lepton number becomes 
lepton parity and the three left-handed neutrinos obtain  small Majorana 
masses through the canonical seesaw mechanism.  It is however also possible 
to keep $B-L$ as a global $U(1)$ symmetry by using a scalar ($\rho$) with 
$L=3$~\cite{mpr13,ms15} instead.

Consider first the simplest self-interacting leptonic dark-matter model 
with the scalars $\chi$ ($L=1$) and $\zeta$ ($L=2$).  If $\rho$ ($L=3)$ 
is used to break $B-L$ spontaneously, then the allowed terms 
$\rho^* \chi^3$, $\zeta^* \chi^2$, and $\rho^* \zeta \chi$ imply that 
a residual $Z_3$ symmetry remains~\cite{mpsz15} as lepton number.  Whereas 
this is sufficient to keep neutrinos as Dirac fermions, $\chi^*$ now 
transforms as $\zeta$, hence the former is no longer stable and cannot be 
dark matter in the present context.  In Ref.~\cite{mpsz15}, although $\zeta$ 
is not stable, it has a very long lifetime because the three $\nu_R$ singlets 
transform as $(-4,-4,5)$ instead of $(-1,-1,-1)$ under $B-L$. 

Assume instead that $\rho$ has $L=4$, then the allowed terms are 
$\rho^* \zeta^2$, $\zeta^* \chi^2$, and $\rho^* \zeta \chi^2$.  This 
results in a residual $Z_4$ symmetry~\cite{hr13,cmsv17} as lepton number, with
\begin{equation}
\nu \sim i, ~~~ \chi \sim i, ~~~ \zeta \sim -1.
\end{equation}
It allows $\chi$ to be self-interacting dark matter.

If $\rho$ has $L=5$, then the terms $\rho^* \zeta^2 \chi$ and 
$\zeta^* \chi^2$ imply a residual $Z_5$ symmetry. 
If $\rho$ has $L=6$, then the terms $\rho^* \zeta^3$ 
and $\zeta^* \chi^2$ imply a residual $Z_6$ symmetry. 
Both would also have $\chi$ as dark matter.
For $L=7$ or greater, only the term $\zeta^* \chi^2$ remains, in which 
case $B-L$ is a global symmetry.

Consider next the self-interacting fermion dark matter $\psi$ with imposed 
dark parity.  The scalar sector now consists of $\zeta$ ($L=2)$ and $\rho$. 
If $\rho$ is assigned $L=3$, then the spontaneous breaking of gauge $B-L$ 
results in a conserved global $B-L$ as desired.

Finally, in the case $L=2$ for $\psi$ as predestined fermion dark matter, 
the scalar sector consists of $\eta$ ($L=4$), $\zeta$ ($L=2)$, and $\rho$. 
If $\rho$ is assigned $L=3$, then the allowed terms are $\eta^* \zeta^2$ 
and $\eta^* \zeta^* \rho^2$.  This means that $\eta^*$ transforms as 
$\zeta$ (so that only one is needed), and the residual symmetry is $Z_6$. 
In that case, 
\begin{equation}
\nu \sim \omega, ~~~ \psi \sim \omega^2, ~~~ \zeta \sim \omega^2,
\end{equation}
where $\omega^6 =1$.  This allows $\psi$ to be self-interacting dark matter.

It is often argued that it is unnatural to have Dirac neutrinos because 
the corresponding Yukawa couplings are so small.  However, a Dirac seesaw 
mecanism~\cite{rs84} works just as well as a Majorana seesaw.  For an explicit 
application, see for example Ref.~\cite{mpsz15}.

\noindent \underline{\it Some Phenomenological Consequences}~:~
In the minimal model of leptonic scalar dark matter, direct detection 
via Higgs exchange is possible, limited only by $\lambda_{01}$ as 
a function of $m_\chi$ from present data.  It is thus always amenable 
to observation in the future.  In the minimal models of leptonic fermion 
dark matter, this effect is supressed in one loop, which means that 
it is not likely to be observable at all.

If the models are supplemented by a gauge $B-L$ symmetry, the resulting 
$Z_{B-L}$ gauge boson is contrained through its production and decay 
at the Large Hadron Collider (LHC) as well as in direct-search experiments.  
The generic requirement~\cite{klq17} is that the lower bound on 
$M_{Z_{B-L}}/g_{B-L}$ increases as $g_{B-L}$ decreases.  For $g_{B-L} = 0.3$, 
it is about 8.7 TeV.  Note that $Z_{B-L}$ does not need to contribute 
to the annihilation cross sections of Eqs.(5) and (10).  However, if 
it does, then it may be revealed in direct-search experiments or at 
the LHC.

Whereas Sommerfeld enhancement may allow $\zeta$ to be produced at present 
from dark matter annihilation, it is difficult to be observed because it 
decays only to neutrinos.  Also, since neutrinos are Dirac fermions, 
neutrinoless double beta decay is forbidden.  A positive such 
signal in future experiments would invalidate the proposed theory.

\noindent \underline{\it Concluding Remarks}~:~
The notion that neutrinos are Dirac fermions leads naturally to the 
conceptual extension of conserved lepton number to dark matter. 
This insight allows for the natural implementation of simple, minimal 
models of self-interacting dark matter with a light scalar dilepton 
mediator which decays only to neutrinos.  As such it solves the problem 
of the possible disruption to the cosmic microwave background caused 
by such a light mediator if it decays to electrons and photons. 
It is simpler and more elegant than the previously proposed 
scenario~\cite{m17-1,m18-1,dsw18} with a gauge $U(1)_D$ symmetry where 
$Z_D$ is the light vector mediator which does not decay, or the very 
recent proposal~\cite{kkyy18} with a gauge $L_\mu-L_\tau$ symmetry.

A byproduct of this investigation is the realization that if gauge $B-L$ 
is the origin of Dirac neutrino masses, its spontaneous breaking (in 
the context of the models being discussed) may result in either global 
$B-L$ or $Z_N$ lepton symmetry.  Indeed, examples of $N=4,5,6$ are 
obtained.  This bolsters the notion that lepton symmetry does not need 
to be either continuous $U(1)$ or odd-even parity, but may in fact be 
somewhere in between, implying also an intimate connection to dark matter.
Instead of $B-L$, the residual $U(1)_\chi$ symmetry in breaking $SO(10)$ 
to $SU(5)$ may also be used.

\noindent \underline{\it Acknowledgement}~:~
This work was supported in part by the U.~S.~Department of Energy Grant 
No. DE-SC0008541.

\baselineskip 16pt
\bibliographystyle{unsrt}

\end{document}